\documentclass[sigconf,screen]{acmart}
\usepackage{diagbox}
\AtBeginDocument{%
  }

\begin{document}

%%
%% The "title" command has an optional parameter,
%% allowing the author to define a "short title" to be used in page headers.
% \title[A User-Centered Approach towards Privacy Protection]{From Privacy Labor to Mediated Privacy: A User-Centered Approach towards Privacy Protection in Online Health Consultations}

\title[From Patient Burdens to User Agency in Online Health Consultations]{From Patient Burdens to User Agency: Designing for Real-Time Protection Support in Online Health Consultations}

%%
%% The "author" command and its associated commands are used to define
%% the authors and their affiliations.
%% Of note is the shared affiliation of the first two authors, and the
%% "authornote" and "authornotemark" commands
%% used to denote shared contribution to the research.
\author{Shuning Zhang}
\orcid{0000-0002-4145-117X}
\email{zsn23@mails.tsinghua.edu.cn}
\affiliation{%
  \institution{Tsinghua University}
  \city{Beijing}
  \country{China}
}

\author{Ying Ma}
\orcid{0000-0001-5413-0132}
\affiliation{%
  \department{School of Computing and \\Information Systems}
  \institution{University of Melbourne}
  \city{Melbourne}
  \country{Australia}}
\email{ying.ma1@student.unimelb.edu.au}

\author{Yongquan `Owen' Hu} 
\orcid{0000-0003-1315-8969} 
\email{yongquan@ahlab.org} 
\affiliation{% 
  \institution{Augmented Human Lab} 
  \institution{National University of Singapore} 
  \country{Singapore} 
}

\author{Ting Dang}
\orcid{0000-0003-3806-1493}
\email{ting.dang@unimelb.edu.au}
\affiliation{%
  \institution{University of Melbourne}
  \city{Melbourne}
  \state{VIC}
  \country{Australia}
}

\author{Hong Jia}
\orcid{0000-0002-6047-4158}
\email{hong.jia@auckland.ac.nz}
\affiliation{%
  \institution{University of Auckland}
  \city{Auckland}
  % \state{Auckland}
  \country{New Zealand}
}

\author{Xin Yi}
\orcid{0000-0001-8041-7962}
\authornote{Corresponding author.}
\email{yixin@tsinghua.edu.cn}
\affiliation{
    \institution{Tsinghua University}
    \city{Beijing}
    \country{China}
}

\author{Hewu Li}
\orcid{0000-0002-6331-6542}
\email{lihewu@cernet.edu.cn}
\affiliation{
    \institution{Tsinghua University}
    \city{Beijing}
    \country{China}
}

%%
%% By default, the full list of authors will be used in the page
%% headers. Often, this list is too long, and will overlap
%% other information printed in the page headers. This command allows
%% the author to define a more concise list
%% of authors' names for this purpose.
\renewcommand{\shortauthors}{Zhang et al.}

%%
%% The abstract is a short summary of the work to be presented in the
%% article.
\begin{abstract}
    Online medical consultation platforms, while convenient, are undermined by significant privacy risks that erode user trust. We first conducted in-depth semi-structured interviews with 12 users to understand their perceptions of security and privacy landscapes on online medical consultation platforms, as well as their practices, challenges and expectation. Our analysis reveals a critical disconnect between users' desires for anonymity and control, and platform realities that offload the responsibility of ``privacy labor''. To bridge this gap, we present SafeShare, an interaction technique that leverages localized LLM to redact consultations in real-time. SafeShare balances utility and privacy through selectively anonymize private information. A technical evaluation of SafeShare's core PII detection module on 3 dataset demonstrates high efficacy, achieving 89.64\% accuracy with Qwen3-4B on IMCS21 dataset. 
  % Online medical consultation is convenient and prevalent in today's society, while its privacy and security concerns persisted with the development of the platform. To address the gap, this paper presents a user-perspective examination of their concerns and privacy protection mechanisms around privacy and security concerns in online medical consultation. Through in-depth semi-structured interviews with users with distinct medical experiences (N=12), we found users input three types of different sensitive information, were most concerned about the deletion rights and user consent, expressed four different expectations regarding their rights, however did not trust the platforms. We further proposed LLM-based technique to facilitate privacy anonymization on online platforms. Our findings are synthesized into three design implications for researchers and clinical workers.
\end{abstract}

%%
%% The code below is generated by the tool at http://dl.acm.org/ccs.cfm.
%% Please copy and paste the code instead of the example below.
%%


%%
%% Keywords. The author(s) should pick words that accurately describe
%% the work being presented. Separate the keywords with commas.
%% A "teaser" image appears between the author and affiliation
%% information and the body of the document, and typically spans the
%% page.
% \begin{teaserfigure}
%   \includegraphics[width=\textwidth]{sampleteaser}
%   \caption{Seattle Mariners at Spring Training, 2010.}
%   \Description{Enjoying the baseball game from the third-base
%   seats. Ichiro Suzuki preparing to bat.}
%   \label{fig:teaser}
% \end{teaserfigure}

\received{20 February 2007}
\received[revised]{12 March 2009}
\received[accepted]{5 June 2009}

%%
%% This command processes the author and affiliation and title
%% information and builds the first part of the formatted document.
\maketitle

\section{Introduction}

Online medical consultation~\cite{al2015online}, a service allowing patients to seek remote advice from healthcare professionals, has seen explosive growth, especially during and after the COVID-19 pandemic. The online consultation market reached RMB 63.62 billion by 2024 globally~\cite{beizhesi}. This prevalent form, typically involving communication via text, audio or video, allows patients to describe symptoms and receive diagnoses or treatment plans from qualified practitioners without a physical visit. 

However, this convenience also comes with severe privacy concerns. Various news have unveiled that online consultation apps has privacy leakage issues~\cite{healthpeople2021,cctv2023}, and patients are concerned about their own privacy leakage~\cite{scol2020,qq2025}. Worse still, some specific patients already experienced the harm caused by the privacy and security harm of online medical consultation~\cite{xinjingbao2024,voc2023}. While these problems are known, a deep, user-centered understanding of how patients perceive these risks and attempt to manage their privacy is underexplored, yet critical for building the trust necessary for sustainable adoption. Therefore, our investigation is guided by the following research questions: (RQ1) What are the perceived privacy risks associated with online medical consultations? (RQ2) What are users' privacy protection practices and expectations? (RQ3) How can a real-time, localized anonymization technique effectively mitigate the aforementioned privacy risks?

To answer these questions, we first conducted an in-depth, semi-structured interview (N=12). We then designed SafeShare, and evaluated its technical implementation using three datasets. For RQ1, we found user apprehensions are not confined to immediate data leakage but extend to the creation of permanent, algorithmic health profiles, which include three primary risk perceptions: (1) platform-level data exploitation, (2) unauthorized disclosure by individual practitioners, and (3) general distrust stemming from profit motives. These concerns are framed by a ``tripartite model of trust'', where users tend to trust the professionalism of individual doctors but remain skepticism towards the platform operators.

For RQ2, we found that users have significant ``privacy labor'' to manage their data, a responsibility offloaded onto them by platforms, including proactive self-censorship such as using pseudonyms, manually cropping identifying details from photos, and blurring information on medical reports. However, they perceive these efforts as both burdensome and ultimately insufficient, leading to ``privacy fatigue''. Therefore, users have clear expectations for system-level reforms, demanding (1) granular control over their data, including the right to be forgotten, (2) enhanced in-context transparency and accountability mechanisms, and (3) robust external regulation, as they do not trust platforms to be self-disciplined.

Finally, for RQ3, we introduce SafeShare, an interaction technique that acts as an intelligent agent within the consultation interface. SafeShare leverages a localized LLM to automate the redaction of Personal Identifiable Information (PII) and provide real-time, context-aware justifications for doctors' data requests, thereby reducing user burden and enhancing transparency. Our technical evaluation shows that the core anonymization module is highly effective, achieving an accuracy of 89.64\% with Qwen3-4B on IMCS21~\cite{chen2023benchmark} dataset in identifying sensitive information in clinical text. Therefore, this paper makes two primary contributions:

% For RQ2, we find that users engage in significant ``privacy labor'' to protect themselves (e.g., manually redacting photos), a process they find both burdensome and ultimately ineffective, leading to ``privacy fatigue''. Consequently, they have clear expectations for platform designs that provide granular control, in-context transparency, and the right to data deletion. 

% Finally, towards RQ3, we present the design of SafeShare, an LLM-based system that automates PII redaction and provides real-time, context-aware justifications for data requests. Our technical evaluation demonstrates that its core technology effectively identifies sensitive information in clinical text with an accuracy of . This paper makes two primary contributions:

$\bullet$ We provide the first formal and in-depth qualitative characterization of users' perceived privacy risks, mitigating strategies, and expectations in online health consultations. 

$\bullet$ We design and evaluate SafeShare, an automatic anonymization technique in medical context capable of running on local devices.

\section{Related Work}

Our work is situated within the broader research landscape of user trust in online healthcare, the privacy dynamics of medical consultation, and the design of privacy-enhancing technologies. A primary challenge in online health services is the establishment of user trust, which is often more fragile in digital settings compared to face-to-face interactions~\cite{mayer2024user}. Trust is a critical mediator influencing a user's willingness to disclose sensitive health information. For example, users show greater willingness to share data with trusted human doctors than with AI systems~\cite{mendel2024advice}. This dynamic frequently forces users into a ``privacy calculus''~\cite{laufer1977privacy}, where they must weigh the diagnostic benefits of sharing data against the perceived privacy risks~\cite{vodicka2013online}. Our paper substantiates this, revealing a clear pattern where users willingly share diagnostic information but strongly resist disclosing personal identifiable information (PII). The perception of control is pivotal in this calculus. When users feel they can manage their own information, their trust and willingness to engage with platforms are enhanced~\cite{luo2023impact}, a sentiment echoed by our participants' demands for granular data control.

A disconnect often exists between the privacy priorities of users and those of technical experts. Users typically focus on control over their personal data, a concern not always shared by expert evaluations that might prioritize technical vulnerabilities~\cite{khoo2024s}. This underscores the need for user-centered privacy solutions. These concerns are not one-sided, as healthcare professionals also report significant anxieties about data security and the trustworthiness of telehealth technologies~\cite{tazi2024we}. Such issues are intensified by the inherent power imbalances in the use of patient-generated health data~\cite{yoo2023discussing}. Indeed, familiarity with a service and the involvement of a trusted care provider are key factors that motivate initial technology adoption~\cite{jung2014virtualized}.

The platforms themselves introduce further complications. Text-based interactions can be perceived as inefficient, prompting physicians to use conversational shortcuts that may compromise the quality of care~\cite{li2023constraints}. While users pose a wide range of questions on these platforms~\cite{ma2018professional}, the underlying app ecosystem is often fraught with privacy risks. Many health apps lack transparent privacy policies~\cite{parker2019private}, are plagued by insecure data handling that enables user profiling~\cite{iwaya2023privacy}, and can cause unintentional harm~\cite{kang2024app}. Analyses of existing privacy policies confirm the urgent need for more robust data protection from both developers and policymakers~\cite{hassan2023unveiling}.

In response, researchers have explored various system-level interventions. Some have studied how user interface design can inform users and address power asymmetries in online consultations~\cite{zhang2024designing}, while others have focused on building tools like inter-doctor recommendation frameworks~\cite{li2025bridging}. Although studies show conversational agents can be as effective as humans for preliminary information gathering~\cite{li2024beyond,li2025comparative}, they do not inherently resolve the privacy dilemma. Our work diverges significantly from these approaches. SafeShare directly confronts the privacy risks prevalent in the ecosystem~\cite{parker2019private,iwaya2023privacy} by  automating the strenuous ``privacy labor'' that we identified, thereby mitigating user burden and enhancing agency.

Other research has focused on privacy redaction. For instance, Zhang et al.~\cite{zhang2024adanonymizer,zhang2024ghost} and Albanese et al.~\cite{albanese2023text} proposed frameworks for automatic text redaction and zero-shot sanitization respectively. However, our work is distinct in its focus on the unique medical context, where we explores how a localized LLM could achieve the balance between privacy and diagnostic utility.

\section{Methodology}

\subsection{Study Setup}

\textbf{Participants and recruitment.} This IRB-approved study recruited 12 participants (5 males, 7 females, age mean=22.2, SD=3.6) through distributing recruiting posters on online chat groups across a week. Participants were required to have at least one time of online consultation experience, however we did not require the total experience of participants as we aimed to recruit participants with diverse symptoms.

\textbf{Interview process and analysis procedure.} We conducted semi-structured interviews with participants to understand their information disclosure behavior on online medical consultation platforms, their cognition towards the information importance and the disclosure risk, their privacy protection behavior, user protection's (in-)effectiveness, the platform's behavior, their perception and the expectations. We conducted all interviews through online Tencent Meeting~\footnote{\url{https://meeting.tencent.com}}. All interviews were recorded and transcribed after acquiring users' consent. We conducted thematic analysis on all interview results: two primary researchers separately coded 20\% of the participants' interview scripts and formed the initial codebook. They then refined the codebook together and iteratively coded the rest of the interview scripts, as the interview was exploratory in nature. We also did not calculate the inter-rate reliability as a criteria because of the exploratory nature of the experiment and according to the previous guidance~\cite{mcdonald2019reliability,horstmann2024those,klostermeyer2024skipping}. We reported the themes in the results section.

\section{RQ1: Information Disclosure, Privacy Risks and Importance}

\subsection{Online Consultation Scenarios: How and Why}

Our analysis analyzes how and why users engage with online medical consultations around two scenarios. 

\textbf{Low-stakes healthcare for triage and access.} Users primarily leveraged online platform as a form of triage for low-acuity medical issues and to gain access to prescriptions. The consultations were often for minor ailments where a physical hospital visit was deemed unnecessary. These included common problems such as \textit{``skin rashes, eye pain''} (P2) and \textit{``colds, fevers, and coughs''} (P1), which were generally perceived as \textit{``not very serious''} (P5). A second key use was prescription fulfillment, particularly for medications requiring a formal consultation step on e-commerce platforms before purchases, such as specific fever reducers or eye drops.

\textbf{A trade-off calculus in platform selection.} Users' decisions to use online or offline services were based on a deliberate weighing of convenience, accuracy and privacy. Online platforms were valued for their immediacy, with users noting, \textit{``I can get a response very quickly''} (P12), a factor especially important for mitigating anxiety. Conversely, for conditions perceived as complex or serious, users defaulted to offline consultations. They values the diagnostic accuracy of physical examinations where a doctor \textit{``can directly touch it and know what kind of nodule it is''} (P3), viewing remote assessments as potentially \textit{``less accurate''} (P3). The anonymity of online platforms was also a significant affordance for sensitive health issues. As P6 noted, for conditions like \textit{``HPV, HIV, it might be more acceptable to consult online''} (P6). However, this was balanced against concerns about data privacy. Some users were wary of platforms retaining their data or requiring them to photograph sensitive areas, preferring the ephemeral nature of an offline visit where \textit{``each consultation is a one-time thing.''} (P2) Finally, the economic model of a platform could also influence the choice. For instance, platforms that limit the number of follow-up questions per payment made offline visits more practical for the extended dialogue required for complex conditions.

\subsection{Information Disclosure and the Privacy Calculus}

We identified two primary findings, the scope of disclosure is highly dependent on the context of the medical interaction, and the privacy calculus that systematically distinguishes between medically necessary data and personally identifiable information. 

\subsubsection{The Scope of Disclosure is Context-Dependent}

Participants did not have a static approach to sharing information but strategically adjusted the breadth and depth of what they disclosed based on the specific goal of consultation. We identified two distinct themes.

\textbf{Formal disclosure for regulated transactions.} When the purpose was to purchase a regulated item, such as a prescription drug, participants recognized the need for formal and verifiable disclosure. They consistently reported providing core identity details, including their name, age, and national ID number, along with their medical history. This was widely viewed not as an invasion of privacy but as a necessary procedural requirement. As one user reasoned, such verification is essential for platforms to prevent the misuse of controlled substances.

\textbf{Holistic disclosure for diagnostic accuracy.} For general diagnostic consultations, the scope of disclosure became significantly broader and more qualitative. To receive an accurate diagnosis, participants shared a holistic view of their condition and lifestyle. This included detailed accounts of \textit{``recent conditions and habits''} (P1), visual evidence such as photos of \textit{``small blisters on the finger''} (P11), and existing \textit{``offline examination reports''} (P12). Participants even divulged seemingly tangential lifestyle details, such as a \textit{``love for frozen or spicy food''} (P3) when consulting for a common cold, operating under the principle that more information would lead to a better diagnosis.

\subsubsection{Privacy Calculus Distinguishes Medical From Personal Data}

Participants performed a consistent mental trade-off to determine what was safe to share. This calculus was defined by two opposing but complementary considerations.

\textbf{Willingness to share diagnostic information.} The first privacy calculus principle is that information perceived as essential for diagnosis is shared willingly, even when acknowledged as private. Participants readily provided detailed symptom descriptions, medical histories, and revealing photographs because the utility of receiving an accurate diagnosis was deemed to outweigh the inherent privacy sensitivity of the data itself. This sentiment was perfectly captured by P5, who stated, \textit{``[My condition] is also private information, [but] I am willing to provide this information to help my diagnosis.''} 

\textbf{Resistance to sharing personal identifiers. }The second and opposing principle is a strong resistance to disclosing PII, especially names and national ID numbers. This information was seen as the critical privacy boundary. The core fear was not the exposure of a medical condition in isolation, but the permanent, verifiable linking of that condition to their real-world identity. One participant powerfully illustrated this fear with the metaphor that connecting their name and ID to their health data \textit{``would be like streaking''} (P11). While the necessity of providing an ID for regulated transactions was sometimes accepted as a form of \textit{``social control''} (P7), the overwhelming consensus was that the ultimate privacy threat lies in the fusion of medical data with personal identifiers.

\subsection{Perceived Privacy Risks}

We identified three themes around the perceived risks associated with the conduct of both digital health platforms and individual medical practitioners. 

\textbf{Perceived risks of platform-level data exploitation.} It captures participants' apprehension about how their information is collected, analyzed, and utilized by the platforms. A primary fear was the unauthorized commercial use of their sensitive health data. One participant articulated the risk of their data being used to \textit{``recommend health supplements and drugs,''} which they believed \textit{``could have an impact on users' health''} (P2). This anxiety was often substantiated by experiences of digital surveillance, with another user noting that after a single online consultation, \textit{``every time I search for something, a pop-up window appears ... I feel like I've been recorded''} (P3). Beyond commercial exploitation, participants feared the creation of permanent and potentially damaging health profiles. For instance, a user expressed alarm that a one-time medication purchase for asthma-like symptoms had resulted in the platform permanently labeling them with an ``asthma'' diagnosis. They worried that \textit{``this file will be shared ... and may affect their ability to purchase insurance in the future''} (P12). The basis for these fears was corroborated by an industry insider who confirmed that platforms systematically analyze user data, stating that \textit{``chat records and communication records can all be heard ... we will use it to do some semantic analysis''} (P11).

\textbf{Anxieties over unauthorized disclosure by individual practitioners.} It relates to the conduct of doctors operating on the platforms. Distinct from the systemic risks posed by the platforms, users hold fear of personal data breaches stemming from individual actions. They were concerned about the potential for unprofessional behavior, as one user worried that a doctor might, \textit{``our of morbid curiosity'',} take a screenshot of their confidential conversation and \textit{``share it with others''} (P11). This highlights a specific vulnerability tied to the perceived integrity and professionalism of the individual doctors.

\textbf{Distrust stemming from profit motives and public data display practices.} It encapsulates a broader skepticism towards the healthcare platforms' business ethics. Participants expressed a general distrust of the commercial incentives driving these services, with one user bluntly stating that \textit{``hospitals and platforms are profit-oriented,''} a concern they felt was especially pronounced with private healthcare entities (P9). This underlying profit motive was seen as a key driver for potential information misuse. Furthermore, this distrust was exacerbated by certain platform features, such as the practice of publishing anonymized medical records for public viewing as case studies. Even with the assistance of anonymization, the practice was a source of significant discomfort. As one participant explained, they would feel violated if their particularly \textit{``outlandish''} case were to be shared publicly.

\section{RQ2: Mitigation and Expectation}

\subsection{User-Initiated Mitigation Methods}

Users described the landscape that, the burden of privacy protection falls largely on the user, with platform-provided measures being perceived as superficial and insufficient. 
% This led to the development of specific user strategies, which exist alongside a deep skepticism of platforms' features. 

\textbf{Proactive self-censorship and information control.} With this most prevalent strategy, users employ tactics such as providing pseudonyms, offering an age range instead of a precise age, or strategically claiming the patient is a \textit{``friend or relative''} to create psychological distance. It also includes the deliberate anonymization of visual data. When submitting images of symptoms, users are diligent about self-censorship, taking care to \textit{``crop out the background''} or ensuring photos are taken against a plain backdrop. One participant stated they would \textit{``definitely erase identifying information like my face or background details.''} (P5) Similarly, when uploading existing medical reports, users would initially \textit{``blur out personal information like their name or ID card number.''}

\textbf{Strategic platform selection based on perceived trustworthiness.} Users gravitate towards services they perceive as reliable and secure. A clear preference was shown for the official applications of large, reputable public hospitals over third-party aggregator platforms. An industry-insider participant justified her exclusive use of a specific hospital's app by noting that on aggregator platforms, \textit{``you have no way to guarantee the quality of the doctors,''} (P3) which she equated with higher privacy risks. This trust is also built on heuristic cues. One user favored a platform that originated from a \textit{``professional medical forum,''} while actively avoiding others with \textit{``chaotic interfaces and promotional ads.''} (P9)

\textbf{Perceived inadequacy of current platform measures.} From the user's perspective, existing platform-level privacy features fail to build meaningful trust. Participants dismissed features like pop-up privacy agreements as superficial formalities that are seldom read or understood. While minor conveniences, such as one platform hiding the medicine name on the delivery package, were noted, they did little to address core data security concerns. The prevailing sentiment is one of profound skepticism. As one user articulated, \textit{``They say they will protect our privacy, but we don't really know if they have.''} (P11)

\subsection{Challenges and Expectations for a Trustworthy System}

The limitations of current measures give rise to significant challenges and a clear set of user expectations for reform. These are rooted in a foundational distrust of platform motives and a desire to reclaim control over their personal health data.

\subsubsection{Core Challenges in Achieving Privacy}

\textbf{Inherent limitations of user-driven mitigation.} While users believe their personal tactics are \textit{``effective to a certain extent,''} they are acutely aware of the limitations. A fundamental challenge is the privacy-utility trade-off: effective diagnosis requires the disclosure of truthful and detailed medical information. As one user noted, \textit{``if you don't say some things, they can't make a diagnosis''} (P6). This necessity often compels disclosure against their better judgment. This is compounded by a sense of learned helplessness, with some users operating under the assumption that their data has \textit{``long been leaked''} through other channels, making extensive protection efforts feel futile.

\textbf{Foundational distrust in the ``black box'' of platform operations.} A significant barrier to trust is the opacity of platform data practices. For users, a platform's internal security mechanisms are a \textit{``black box''} (P11), making it impossible to verify security claims. This distrust is structured around a tripartite model of trust: users generally trust doctors' professionalism (\textit{``doctors have medical ethics''} (P10)), but this trust does not extend to the platforms, which are viewed as profit-driven entities. One user expressed this dichotomy starkly: \textit{``I trust their professional ability, but I default to assuming my information is being leaked [by the platform]''} (P12). This fear was validated by an insider who confirmed that platforms use consultation data to create a \textit{``semantic database for analysis''} to improve their products, a practice described as \textit{``dancing on a dangerous edge''} (P11).

\subsubsection{User Expectations for System-Level Change}

Fueled by these challenges, users articulated a clear vision for a trustworthy system, centered on demands for technical control and regulatory oversight.

% A primary user demand is for direct and granular control over their data.
\textbf{Demand for granular user control and data ephemerality.} This includes the ability to remain anonymous to the practitioner, with one participant wishing for a system where \textit{``the doctor cannot see my personal data.''} Users want the power to \textit{``freely choose which parts of the medical record to display to the doctor.''} A key component of control is the \textit{``right to be forgotten.''} Users desire the assurance that \textit{``after the consultation ends, my personal information can be erased from their system''} (P2), viewing this as far more meaningful than a simple promise not to misuse data. The ideal, for some, is a system that severs the link to a permanent identity entirely, where \textit{``I can just pay for the service without needing to log in or bind my phone number''} (P12).

% Instead of ambiguous legal policies, users call for explicit, in-context guarantees.
\textbf{Demand for enhanced transparency and accountability mechanisms} A participant suggested the chat interface should feature a prominent \textit{``claim stating that your chat history will only be used for a specific purpose and that screenshots or recordings are prohibited''} (P11). To enforce such policies, users expect platforms to implement \textit{``very strict privacy training and security tests''} for doctors and to establish robust complaint channels and evaluation systems to hold both practitioners and platforms accountable for their conduct.

% There is a pervasive belief that platforms cannot be relied upon to regulate themselves.
\textbf{An imperative for external governance and regulation.} Users expressed a profound lack of faith in corporate self-discipline and a corresponding demand for external oversight. This sentiment was captured unequivocally by one participant: \textit{``I think the platform will never be self-disciplined ... only strong external constraints, like national policies, would make me feel at ease''} (P12)

\section{SafeShare: A Data Protection Interactive Technique}

To address the privacy tensions identified in our study, we designed SafeShare, an interactive technique that reframes data protection not as simple redaction, but as a process of contextual anonymization. This approach is designed around balancing the disclosure of medically relevant information with the robust protection of PII. SafeShare functions as an intelligent intermediary within the chat interface, leveraging a localized LLM to empower users with both the tools and the understanding to navigate the trade-off, thereby enhancing user agency and trust.

\begin{figure}[!htbp]
    \includegraphics[width=0.5\textwidth]{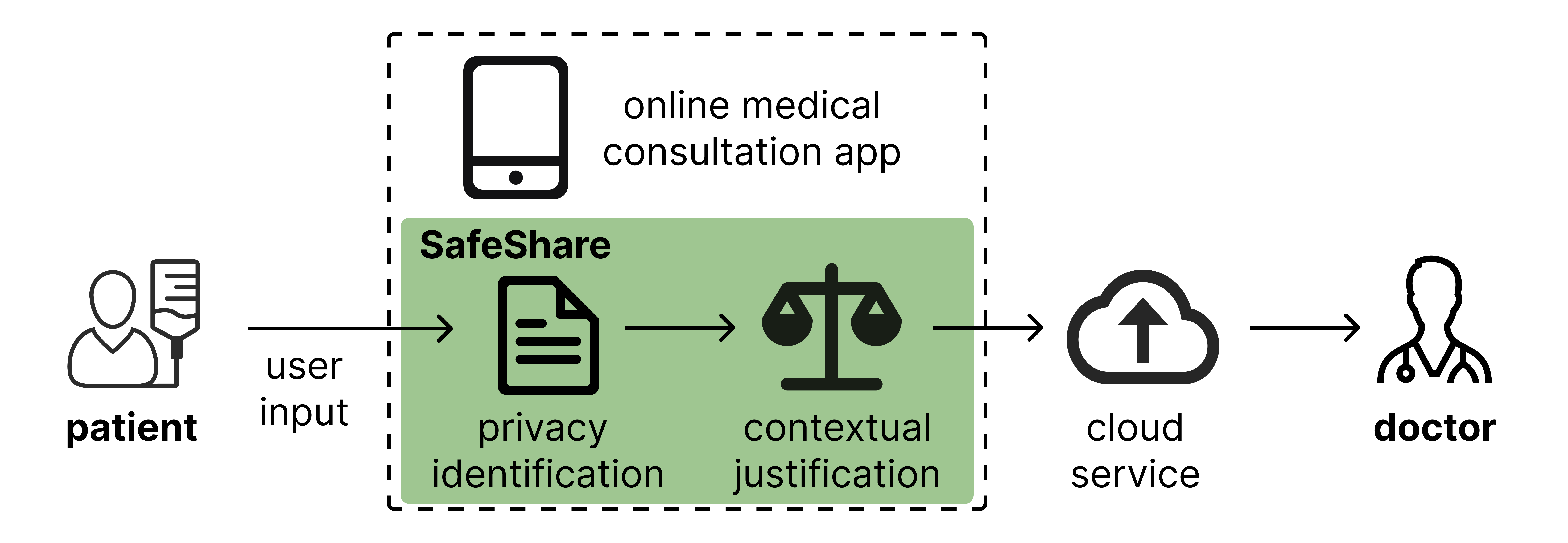}
    \caption{SafeShare acts as a bridge between users and cloud service, anonymizing medical private information.}
\end{figure}

SafeShare comprises of a real-time anonymization and an in-context justification module. The real-time anonymization module automatically detects potential PII in both text and images according to pre-defined categories~\cite{zhou2025rescriber} elaborated in the prompts. Then the justification module, upon receiving the full set of identified entities, analyzes the user's query history to discern its specific diagnostic or information intent. Based on this intent, the module dynamically determines which identified entities are important for answering the query and which are not. This mechanism allows SafeShare to generate an optimal anonymization list for each specific context, describing which sensitive information to anonymize. It directly automates the strenuous and error-prone ``privacy labor'' and alleviating the cognitive load on the user.
% dynamically decides which sensitive entities to anonymize with the specific question, among all detected sensitive entities, balancing privacy and diagnosing utility. This process is designed to directly automate the strenuous ``privacy labor'' and cognitive load of users.  

After selectively anonymizing the inputs, SafeShare parses the output and replaces the users' original input locally, thereby protecting users' medical private information.

\section{Evaluation of SafeShare}

To validate the feasibility SafeShare, we conducted a quantitative evaluation, focusing on the performance of its core component: the real-time anonymization module. The objective was to assess the module's accuracy in identifying diverse types of PII from realistic medical text, which is critical for its ability to automate the ``privacy labor'' identified the interview.

\subsection{Experimental Setup}

\textbf{Dataset.} We utilized three different datasets related to online consultation. MedDG~\cite{liu2022meddg} is a large-scale, entity-centric Chinese medical dialogue dataset collected from Chinese online consultation platform \textit{Doctor.Chunyu}\footnote{\url{https://www.chunyuyisheng.com}}, with 17,864 Chinese dialogues, 385,951 utterances. ReMeDi~\cite{yan2022remedi} contains 96,965 conversations and 843 types of diseases, including 1,557 conversations with fine-gained labels from \textit{Doctor.Chunyu}\footnote{\url{https://www.chunyuyisheng.com}}. IMCS21~\cite{chen2023benchmark} contains 4,116 samples, 10 diseases with 164,731 utterances from a Chinese online community, Muzhi\footnote{\url{http://muzhi.baidu.com}}. 
% UTHealth Shared Task dataset~\cite{}, a standard and wide-used benchmark for the de-identification of clinical notes. This dataset contains expert-annotated clinical records featuring 25 distinct PII tags across seven primary categories (e.g., \texttt{patient}, \texttt{doctor}, \texttt{hospital}, \texttt{location}, \texttt{date}, \texttt{id}, \texttt{contact}). Its use of real-world clinical narratives makes it an appropriate corpus for validating a system intended for deployment in medical consultation contexts.

\textbf{Models and Metrics.} We selected leading models with different sizes, brands and open or closed source status, including GPT-4o-mini from OpenAI, Deepseek-R1-7B from Deepseek, different model sizes of Qwen3 from Alibaba. We selected models with parameters smaller than 8B for evaluating on-device anonymization performance. The evaluation was conducted in a zero-shot setting, where each model was given a structured prompt defining the PII categories (e.g., NAME, ID) and instructed to extract all corresponding entities from the input text, balancing latency, cost and accuracy. The detailed anonymization prompts are shown in Appendix~\ref{app:prompt}.

\subsection{Performance and Illustrative Case}

\noindent\textbf{Quantitative Results.} The quantitative performance of the selected LLMs was evaluated on IMSC21, MedDG and ReMeDi. The evaluation centered on both anonymization accuracy and appropriateness (see Appendix~\ref{app:metrics} for detailed definition and calculation of these metrics), which separately quantified the capability of correctly identifying sensitive information, and preserves the necessary clinical information for diagnosis. Table~\ref{tab:llm_results} and~\ref{tab:llm_appropriate} showed the results. The findings indicated a clear performance trade-off among the models. \textit{Qwen3-4b} model showed the highest anonymization accuracy across all three datasets, achieving scores of 89.64\%, 84.86\% and 82.41\%, however exhibited lower scores in anonymization appropriateness. 

\begin{table}[!htbp]
\centering
\caption{Anonymization accuracy of LLMs on different medical datasets.}
\label{tab:llm_results}
\begin{tabular}{lccc}
\toprule
\diagbox{Model}{Dataset} & IMCS21 & MedDG & ReMeDi \\
\midrule
DeepSeek-R1-7B & 77.96\% & 70.00\% & 75.52\% \\
GPT-4o-mini & 78.57\% & 72.77\% & 74.61\% \\
Qwen3-1.7B & 78.01\% & 73.22\% & 75.82\% \\
Qwen3-4B & 89.64\% & 84.86\% & 82.41\% \\
Qwen3-8B & 78.70\% & 87.04\% & 71.27\% \\
\bottomrule
\end{tabular}
\end{table}

Conversely, \textit{Qwen3-1.7B} model showed the highest scores for anonymization appropriateness, with 92.40, 87.97 and 95.28 on IMCS21, MedDG, and Medical Dialogue datasets respectively out of 100. This suggests that while it was less precise in PII removal, it better preserves the diagnostic utility of the clinical text. Other models like \textit{DeepSeek-R1-7B} and \textit{GPT-4o-mini} achieved balanced performance, underscoring the potential of LLMs for balancing privacy and utility.

\begin{table}[!htbp]
\centering
\caption{Appropriateness of LLM's anonymization for diagnosing the symptom.}
\label{tab:llm_appropriate}
\begin{tabular}{lccc}
\toprule
\diagbox{Model}{Dataset} & IMCS21 & MedDG & ReMeDi \\
\midrule
DeepSeek-R1-7B & 80.21 & 91.03 & 78.49 \\
GPT-4o-mini & 75.65 & 89.98 & 70.62 \\
Qwen3-1.7B & 92.40 & 87.97 & 95.28 \\
Qwen3-4B & 78.91 & 80.36 & 73.90 \\
Qwen3-8B & 70.04 & 76.03 & 68.44 \\
\bottomrule
\end{tabular}
\end{table}

\noindent\textbf{Illustrative Case.} To demonstrate the module's practical application, we presented a case of SafeShare, with inputs and output anonymization result, which contains multiple PII types\footnote{The original information is replaced for anonymization.}:

\begin{quote}
\textit{Original User Input:} ``I am worried about the test results for my daughter Jane Doe from her appointment on May 20, 2025, with Dr. Smith at Peking University Hospital. We can be reached at 138-0000-0000 if needed.''
\end{quote}

SafeShare would present the following redacted version to the user for one-click approval before transmission:

\begin{quote}
\textit{SafeShare Anonymized Output:} ``I am worried about the test results for my daughter [PATIENT] from her appointment on May [DATE], 2025, with [DOCTOR] at [HOSPITAL]. We can be reached at [PHONE] if needed.''
\end{quote}

This case illustrates the module's ability to correctly identify and appropriately anonymize PII entities within patients' description. SafeShare anonymizes sensitive information while retaining the meaning for the doctors to understand and diagnose.
% By automating this process, the system effectively reduces the cognitive load on the user and provides a reliable mechanism for privacy protection, directly addressing the core issues identified in our study.

\section{Discussions}

% Our analysis reveals three central privacy tensions within online medical consultations. The first is the conflict between regulatory compliance, which often requires user identification, and the user's strong desire for anonymity. The second is the significant cognitive burden of ``privacy labor'', as users are made responsible for their own data protection. The third involves new data portability risks that emerge as users construct hybrid online-offline health journeys. These themes collectively expose a critical gap between user expectations for privacy and the reality of current platform designs.

% This section embeds the problems of online medical consultation in the broad socio-technical context, first discussing regulation compliance, then its feasibility and applicability, finally the culture nuance and adaptations.

\textbf{Regulation compliance and anonymity.} A core tension emerges between platform-enforced identification protocols and users' desire for anonymity. While some users understand the need for compliance, such as providing a real ID for purchasing controlled prescription drugs, as a necessary trade-off for societal safety, the general preference is for anonymous interaction~\cite{ahn2020balancing}. This desire is particularly strong when dealing with stigmatized conditions where anonymity feels like a prerequisite for seeking care. However, users' attempts to maintain privacy through tactics like providing false information are often thwarted by system design. For instance, platforms can enforce compliance through mandatory real-name verification or by rejecting invalid ID numbers. 
% This leaves users in a position of powerlessness, compelling them to sacrifice anonymity or forgo care.
% , highlighting the need for systems that can better navigate this conflict, perhaps by offering tiered anonymity based on the service requested.

% SafeShare could be integrated into commercial platforms featuring transparent communication~\cite{shen2024systematic}. Besides detecting and anonymizing, SafeShare could also transparently communicate to users about the potential privacy risks~\cite{lee2024priviaware}. 

% Our findings also suggest that platforms largely delegate the responsibility for privacy protection to their users. Participants described engaging in constant and effortful ``privacy labor,'' such as carefully redacting personal details from lab reports, cropping photos to remove identifying backgrounds, and continuously vetting the necessity of a doctor's questions. This places a significant cognitive load on patients who are often already in a state of anxiety about their health. Furthermore, this model of user-managed privacy is inherently unreliable and unsustainable. We observed clear instances of ``privacy fatigue,''~\cite{choi2018role} where a user who was initially diligent about redacting her documents eventually stopped due to the effort involved. This offloading of responsibility is a design failure that results in inconsistent protection. An effective approach would embed robust and usable privacy measures into the system by default, shifting the burden from the user to the platform designer.

\textbf{Feasibility of SafeShare.} Our findings indicate a prevalent design failure in which communication platforms delegate the onus of privacy protection to users. This delegation compels individuals to engage in constant and effortful ``privacy labor,'' such as manually redacting personal details from documents, cropping identifying features from images, and continuously assessing the necessity of information requests. Such practices impose a significant cognitive load on users, who may already be in a state of vulnerability. This model of user-managed privacy is not only burdensome but also inherently unreliable and unsustainable. We observed clear instances of ``privacy fatigue''~\cite{choi2018role}, where users ceased their redaction efforts due to the sheer effort involved. 
% The consequence of this offloaded responsibility is inconsistent and often inadequate privacy protection.

To address these shortcomings, an effective approach would embed robust and usable privacy measures directly into the technique's design, thereby shifting the primary responsibility from the user to the platform. A technique like SafeShare could be integrated into commercial platforms that feature transparent communication channels~\cite{shen2024systematic}. Beyond merely detecting and anonymizing sensitive information, such a technique could also transparently communicate potential privacy risks to users~\cite{lee2024priviaware,zhang2025privcaptcha}, therefore creating a supportive and secure environment.

\textbf{Online, offline consultation and privacy risks.} User privacy preferences and disclosure behaviors are highly contingent upon the specific medical condition~\cite{li2024beyond}. Concerns range from the mandatory reporting of travel history for infectious diseases~\cite{liu2022privacy} and the fear of social stigma associated with conditions like HIV~\cite{bussone2016disclose} and mental health issues~\cite{hu2024exploring}, to a reluctance to share basic personal identifiers for common ailments~\cite{liu2024prevalence}. This practice introduces a critical privacy challenge centered on data portability, as sensitive information is transferred from a trusted, regulated healthcare environment to a less secure digital one. This transition represents a significant point of vulnerability, underscoring the need for platform design that accounts for the broad, multi-modal health ecosystem.

\begin{acks}
This work was supported by the Natural Science Foundation of China under Grant No. 62472243 and 62132010.
\end{acks}

%%
%% The next two lines define the bibliography style to be used, and
%% the bibliography file.
\bibliographystyle{ACM-Reference-Format}
\bibliography{sample-base}

%%
%% If your work has an appendix, this is the place to put it.
\appendix

\section{Prompt Structure}\label{app:prompt}

We used LLM-powered anonymization because it could potentially understand the task, ensuring both privacy and utility. Besides, ensuring localized processing could minimize the data uploading, thereby protecting privacy and mitigating privacy risks. In the next subsections, we provided the design and implementation of the prompts, including the anonymization prompt and evaluation prompt.

\subsection{Anonymization Prompt}

The set of entity categories targeted for anonymization was established in accordance with prior privacy frameworks, encompassing: name, email, phone number, ID, online identity, geolocation, affiliation, demographic attributes, time, financial information, and educational records~\cite{zhou2025rescriber}. Different from prior work~\cite{zhou2025rescriber}, to execute the NER task using a LLM, balancing diagnosis and anonymization, we engineered a structured prompt to precisely guide its behavior.

The prompt's architecture is multifaceted. First, it employs a role-playing instruction, assigning the model the persona of a \textit{professional medical and privacy expert specializing in NER} to contextualize the task. Second, it defines the primary objective, which extends beyond simple entity extraction to enable a deliberate balance between privacy preservation and data utility. The model is tasked with identifying and extracting textual instances of sensitive entities whose sensitivity overweigh utility, from a given medical dialogue (input as \textit{\}dialogue\_text\}}) based on a predefined set of categories (input as \textit{\{entity\_list\_str\}}). This identification is a prerequisite for subsequent anonymization, ensuring that non-sensitive, clinically relevant information remains intact, thereby preserving the utility of the data.

Third, it imposes a strict, machine-readable output schema that is critical for automated downstream processing. The prompt mandates a valid JSON output where keys correspond to the entity categories and values are lists of the exact textual excerpts. This structural requirement is reinforced with a one-shot example and includes instructions for handling null results (i.e., using an empty list or omitting the key) to ensure consistent and predictable model behavior.

\subsection{Evaluation Prompt}

To evaluate the dual objectives of effective PII removal and the preservation of clinical utility, we meticulously designed two prompt. The designs were necessary to operationalize the measurement of our two primary metrics, anonymization accuracy and anonymization appropriateness, while preventing task contamination and ensuring the validity of each measure. This approach allows for a focused and unbiased assessment of each objective independently.

The first prompt is designed exclusively to assess anonymization accuracy. To this end, we assign the LLM the persona of a privacy expert. The prompt provides the model with both the original, unaltered dialogue and the list of PII entities extracted by our system. The LLM is then tasked with a technical validation: to evaluate the correctness and completeness of the extracted entities against the source text one by one. We aggregated each assessment after using LLMs' evaluation.

The second prompt is engineered to evaluate anonymization appropriateness. Here, the LLM is assigned the different persona of a clinical physician. Critically, it is provided only with the fully anonymized version of the dialogue, with no access to the original text or the redacted PII. The model's task is to determine whether the remaining clinical information is sufficient to make a meaningful medical diagnosis, through providing a quantitative score.

The rationale for this two-prompt design is to create a controlled evaluation environment. By isolating the tasks, we ensure that the LLM's assessment of diagnostic utility (Prompt 2) is not biased by its knowledge of what specific PII was removed (Prompt 1). This separation is crucial for obtaining a reliable and objective measure of the delicate balance between privacy protection and data utility, providing a scalable and methodologically sound alternative to using human experts for each distinct evaluation task.

% Second, it explicitly defines the objective: to extract all privacy textual instances of entities from a given medical dialogue (input as \textit{\{dialogue_text\}}) that correspond to a dynamically provided list of categories (input as \textit{\{entity_list_str\}}).

% Third and critically for downstream processing, the prompt imposes a strict output schema. It mandates that the model's response be a valid JSON object. Within this object, the keys must be the predefined entity categories, and their corresponding values must be lists of strings, where each string is an exact quotation of an entity identified in the text. This enforcement ensures the output is consistently structured and machine-readable. To handle null cases robustly, the instructions specify that any category for which no entities are found must be represented by an empty list or be omitted from the JSON object entirely. Finally, the prompt includes a one-shot example to provide a concrete illustration of the expected input-to-output transformation, minimizing ambiguity and improving model compliance.

\section{Evaluation Metrics}\label{app:metrics}

To validate the effectiveness of the anonymization process, we employed two metrics: anonymization accuracy and anonymization appropriateness. Anonymization accuracy refers to the correctness of redacting private information, while anonymization appropriateness evaluates whether the anonymized answer retains sufficient information for an accurate patient diagnosis.

We estimated anonymization accuracy by using an advanced LLM (qwen-max) to judge the correctness of the entity recognition. To estimate anonymization appropriateness, the LLM was prompted to role-play as a physician and determine if it could accurately discern the patient's symptoms from the anonymized text.

Notably, to validate the accuracy metric, one experimenter manually coded a sample of 50 records. The inter-rater reliability between the manual coding and the LLM's judgments, calculated using Cohen's Kappa~\cite{kvaalseth1989note}, was 0.81, indicating substantial agreement. We decided to use an advanced LLM instead of human annotators because the dataset contains a wide variety of symptoms and disease categories, making it infeasibly complex to recruit physicians who are expert across all these different areas. We also acknowledge that future work could benefit from having human physicians perform the annotation task.

\end{document}